\documentclass{aa}  
\usepackage{graphicx}
\usepackage{txfonts}
\begin{document}
   \title{Photometry of GSC~762-110, a new triple-mode radially pulsating star
\thanks{Tables~2-4 are only available in electronic form at the CDS via anonymous ftp to cdsarc.u-strasbg.fr (130.79.128.5) 
or via http://cdsweb.u-strasbg.fr/cgi-bin/qcat?J/A+A/.}
}

   \author{Patrick Wils\inst{1}
           \and
           Ioannis Rozakis\inst{2, 3}
           \and
           Stelios Kleidis\inst{2, 3}
           \and
           Franz-Josef Hambsch\inst{4, 1}
           \and
           Klaus Bernhard\inst{5, 4}
          }

   \institute{Vereniging Voor Sterrenkunde, Belgium;
              \email{patrickwils@yahoo.com}
         \and
              Zagori Observatory, Epirus, Greece;
              \email{astrozakis@yahoo.com; steliosklidis@gmail.com}
         \and
              Helliniki Astronomiki Enosi, Athens, Greece
         \and
             Bundesdeutsche Arbeitsge\-mein\-schaft f\"{u}r Ver\"{a}nderliche Sterne e.V. (BAV), Germany;
             \email{hambsch@telenet.be}
         \and
             A--4030 Linz, Austria;
             \email{klaus.bernhard@liwest.at}
             }

   \date{Received ?? ??, 2007; accepted ?? ??, 2007}

 
  \abstract
  {Stars pulsating in three radial modes are very rare, only three examples are known in the Galaxy.
These stars are very useful since their periods may be measured very precisely, and this will constrain the global stellar parameters 
and the models of the star's interior.}
   {The purpose of this paper is to present a new example of the class of triple-mode radial pulsators.}
   {A search for candidate multi-mode pulsators was done in public survey data.  
    Time-series photometry of one of the candidates, \object{GSC~762-110}, was performed.}
   {GSC~762-110 was found to be a triple-mode radial pulsator, with a fundamental period of 0.1945d and period ratios of 0.7641 and 0.8012.
    In addition two non-radial modes were found, for which the amplitude diminished considerably over the last few years.
   }
   {}


   \maketitle
%

\section{Introduction}

Multi-mode radially pulsating stars are important in the theory of pulsating stars.
Pulsation models allow to precisely determine their physical parameters.
Double-mode radial pulsation is observed quite frequently in high amplitude
$\delta$ Scuti (HADS) and SX~Phe stars, in RR~Lyrae variables and in classical Cepheids.
In most cases the fundamental and first overtone mode are excited, but a few Cepheids 
are known to pulsate in the first and second overtone mode (Beltrame \& Poretti, \cite{beltrame}).
Triple-mode pulsation is however very rare.  Only three stars in the Galaxy are known
to pulsate in the first three radial modes (the fundamental mode plus the first and second overtone): 
\object{AC~And} (Kov\'acs \& Buchler, \cite{acand}), \object{V829~Aql} (Handler et al., \cite{v829aql}) and \object{V823~Cas} (Jurcsik et al., \cite{v823cas}).
Despite their long or fairly long fundamental periods, more typical for the RR~Lyrae regime, they are all considered to be evolved $\delta$ Scuti stars.
In addition two short period Cepheids in the Large Magellanic Cloud were found to pulsate 
in the first three overtone modes (Moskalik \& Dziembowski, \cite{lmc}).  
Because of their importance for stellar models, more examples of radial triple-mode pulsators are welcome.

Public survey data from the All Sky Automated Survey (ASAS-3; Pojmanski, \cite{asas}) was searched for possible new multi-mode radial pulsators.  
The results on double-mode RR Lyrae variables were published by Bernhard \& Wils (\cite{rrd}) and references therein.
\object{ASAS~071220+0921.1} = \object{GSC~762-110} at 
position $\alpha_{2000}= 07^{h}12^{m}19\fs412$, $\delta_{2000}= +09\degr21\arcmin02\farcs66$ (Hog et al., \cite{tycho2}), 
was found to be a very good candidate for a triple-mode pulsator.
Follow-up photometry was therefore initiated at four private observatories.
The results of this campaign are discussed in the following sections.


\section{Observations and Frequency Analysis}

\begin{figure*}
\centering
\includegraphics[width=17cm]{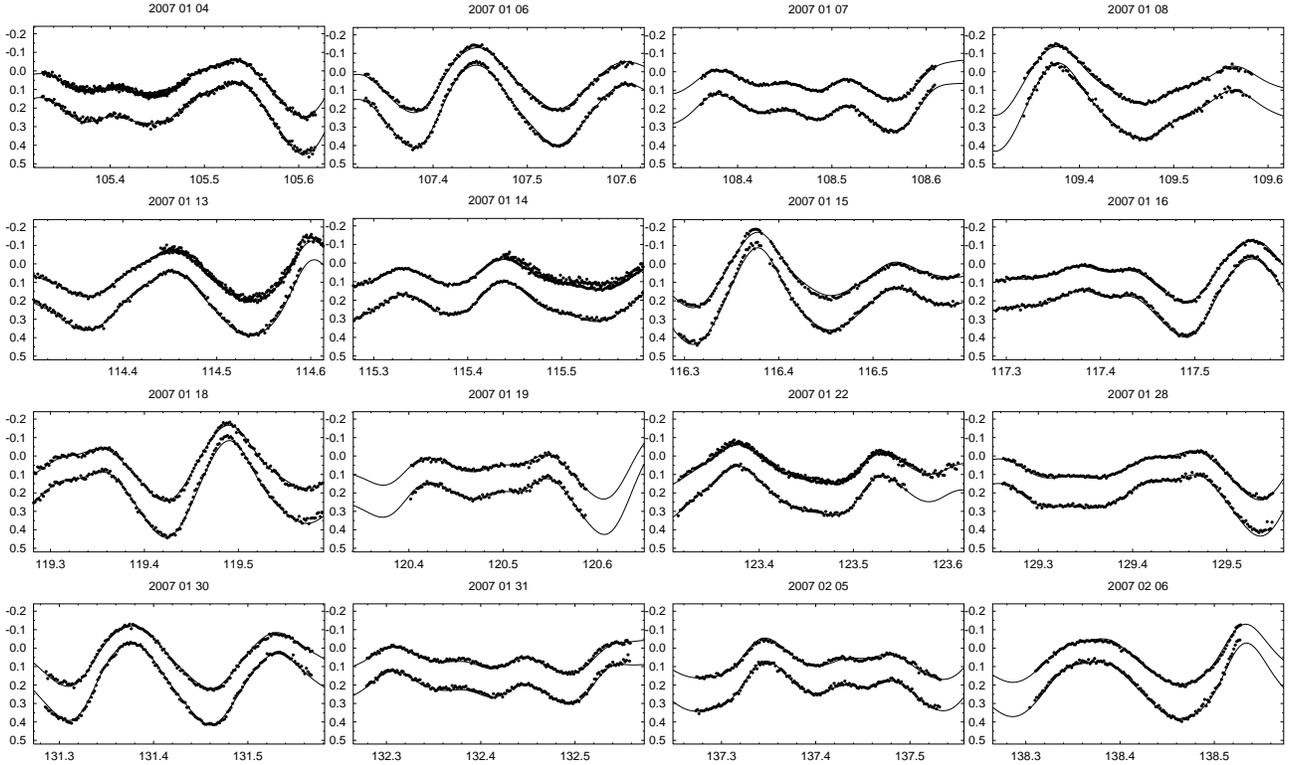}
\caption{Differential $B$ (lower curve) and $V$ (upper curve) data of GSC~762-110 on a number of nights.  
Labels on the horizontal axis are HJD~-~2454000.
The model plots from Table~\ref{freq} are shown as well.}
\label{plot}
\end{figure*}

GSC~762-110 was observed on 68 nights in a five month period during the 2006/2007 season.  
More than 300 hours of CCD photometry were secured, mostly in $V$, but also 150 hours in $B$ and 50 hours in $I_C$.
The complete observation log, together with the instruments used, is given in Table~\ref{log}.  
Differential photometry was performed using \object{GSC~766-2426} (a star of spectral type A2 according to the SAO star catalogue) as comparison star and \object{GSC~762-16} or \object{GSC~766-922} as check stars.
Nightly standard deviations on the difference between the check star and comparison star were usually better than 0.01 magnitude.
No instrumental corrections were applied on the data.
Some nightly plots are shown in Fig.~\ref{plot}.
The photometric data are available electronically at the CDS.
In Tables~2-4, Column~1 lists the HJD of the observations, and Column~2 gives the differential magnitudes of 
GSC~762-110 with respect to GSC~766-2426 for the $B$, $V$, and $I_C$ colours.

From a $B-V$ value of 0.294 for GSC~766-2426 (ESA, \cite{tycho}), an average $B-V$ of
0.44 has been derived for GSC~762-110.

\begin{table*}
\begin{center}
\caption{Observation log.}
\label{log}
\begin{tabular}{clllccrccc}
\hline
Observer & Telescope & CCD camera & Timespan & Nbr of & Nbr of & \multicolumn{3}{c}{Number of data points} \\
initials &           &            &  {\small (JD-2450000)} &  nights & hours & $B$ & $V$ & $I_C$ \\
\hline
KB    & 20-cm C8    & SX Starlight  & 4083-4189 & 14 &  49.0 & -    & 1960 & - \\
HMB   & 35-cm C14   & SBIG ST-8     & 4066-4186 & 16 &  51.6 & -    & 1039 & 446 \\
SK    & 30-cm LX200 & SBIG ST-7XMEI & 4105-4200 & 29 & 150.1 & 3649 & 3629 & - \\
JR    & 35-cm C14   & SBIG ST-7XMEI & 4068-4175 & 18 &  78.4 & -    & 4391 & - \\
Total ($V$) &             &         & 4066-4200 & 68 & 303.0 & -    & 11019 & - \\
\hline
\end{tabular}
\end{center}
\end{table*}

Period04 (Lenz \& Breger, \cite{period04}) was used for the frequency analysis.  
The ASAS-3 data set ($V$ band photometry) and our own data set were treated independently.  
Table~\ref{freqasas} lists the frequencies detected in the ASAS-3 $V$ data with their semi-amplitude in millimag and the signal to noise ratio ($S/N$). 
The latter was calculated in a box with a size of 20\% of the frequency, centered on the frequency itself.
Only frequencies with $S/N > 4$ were retained (Breger et al., \cite{breger}). 
Uncertainties on the frequency and the amplitudes are given between parenthesis in units of the last specified digit. 
The uncertainties were calculated using the Monte-Carlo simulations provided by Period04.

Five independent frequencies were detected. 
The first three modes $f_0$, $f_1$ and $f_2$ have fairly large amplitude and, as will be seen further, only a small
phase difference in $B$ and $V$.  Furthermore their ratios agree with the theoretically calculated ratios
for the first three radial modes of $\delta$ Scuti stars (Fitch, \cite{fitch}).  This means that $f_0$, $f_1$ and $f_2$ are almost certainly the fundamental, and the first and second overtone radial mode.  
Two additional modes $f_3$ and $f_4$, likely non-radial, were also detected, both at a higher frequency than the second overtone mode.
In addition, the first harmonic of the fundamental radial mode, and two combination frequencies of the radial modes were found.

Table~\ref{freq} lists the frequencies detected in our own data.
In addition to the semi-amplitude and signal to noise ratio, the phase of the sine term in the Fourier series (given with respect to $t_0$ = HJD~2450000) 
for the frequencies in $B$ and $V$ are also listed.   
The last two columns in Table~\ref{freq} give the amplitude ratio and phase difference between the $B$ and $V$ frequencies.
Model plots based on these frequencies were added in Fig.~\ref{plot}.

The same independent frequencies as in the ASAS-3 data set were found in our data sets, 
and because of the higher precision of our data many more combination frequencies could be resolved.
Although the $V$ data set is the largest and has the longest timespan, 
not all combination frequencies that were found in $B$ were detected in $V$, because of their smaller amplitudes.
In particular also the non-radial mode $f_4$ was detected only in $B$ (see the next section for further discussion).
Note that the combination frequency $2f_0+f_1$ could not be distinguished from $f_2+f_3$, 
so that it is not clear which one of those two was found to be excited.

The three radial modes, the non-radial mode $f_3$ and a number of combination frequencies were found in the $I_C$ data as well.  
However, because of the smaller number of data points and nights (only ten), the smaller amplitudes and the smaller timespan of the observations, 
the uncertainties on the amplitude and phase values were about an order of magnitude larger than in $B$ and $V$.  
These results were therefore not included in the Table~\ref{freq}.

\setcounter{table}{4} 
\begin{table}
\begin{center}
\caption{Frequencies for GSC~762-110 found in ASAS-3 data.}
\label{freqasas}
\begin{tabular}{cr@{}lrr@{}lrr}
\hline
\multicolumn{3}{c}{Frequency} & \multicolumn{1}{c}{$A_V$} & $S/N$ \\
 & \multicolumn{2}{c}{(c/d)}  & \multicolumn{1}{c}{(mmag)} \\
\hline
$f_0$ & 5.14124 & (1) & 77(3) & 29.6 \\
$f_1$ & 6.72845 & (1) & 74(2) & 28.4 \\
$f_2$ & 8.39755 & (2) & 45(2) & 16.9 \\
$f_3$ & 8.60796 & (4) & 21(3) & 8.1 \\
$f_0+f_1$ & 11.86970 & & 18(2) & 6.7 \\
$f_4$ & 8.96428 & (7) & 13(2) & 4.8 \\
$2f_0$ & 10.28249 & & 16(2) & 5.8 \\
$f_2-f_1$ & 1.66910 & & 13(3) & 4.9 \\
\hline
\end{tabular}
\end{center}
\end{table}

\begin{table*}
\begin{center}
\caption{Frequencies for GSC~762-110 detected in our data sets. }
\label{freq}
\begin{tabular}{cr@{}lrr@{}lrrr@{}lrrr@{}lrlr@{}ll}
\hline
\multicolumn{3}{c}{Frequency} & \multicolumn{1}{c}{$A_B$} & \multicolumn{2}{c}{$\Phi_B$} & \multicolumn{1}{c}{S/N} 
                              & \multicolumn{1}{c}{$A_V$} & \multicolumn{2}{c}{$\Phi_V$} & \multicolumn{1}{c}{S/N} 
                              & \multicolumn{1}{c}{$A_B/A_V$} & \multicolumn{2}{c}{$\Phi_B$ - $\Phi_V$} & \\
 & \multicolumn{2}{c}{(c/d)}  & \multicolumn{1}{c}{(mmag)} & \multicolumn{2}{c}{(degrees)} & \multicolumn{1}{c}{$B$}
                              & \multicolumn{1}{c}{(mmag)} & \multicolumn{2}{c}{(degrees)} & \multicolumn{1}{c}{$V$}
                              &           & \multicolumn{2}{c}{(degrees)} & \\
\hline
$f_0$ & 5.1412 & (3) & 95.9(3) & 256.5 & $\pm$ 0.2 & 116.4 & 74.6(3) & 256.1 & $\pm$ 0.2 & 72.3 & 1.286(7) & 0.5 & $\pm$ 0.3 &  \\
$f_1$ & 6.7284 & (3) & 94.1(3) & 75.6 & $\pm$ 0.2 & 139.6 & 74.3(4) & 76.2 & $\pm$ 0.3 & 88.8 & 1.266(8) & -0.7 & $\pm$ 0.3 &  \\
$f_2$ & 8.3974 & (4) & 48.6(4) & 328.6 & $\pm$ 0.4 & 79.6 & 39.5(3) & 329.8 & $\pm$ 0.5 & 50.2 & 1.230(14) & -1.2 & $\pm$ 0.6 &  \\
$f_0+f_1$ & 11.8695 & & 18.6(4) & 139.1 & $\pm$ 1.0 & 36.3 & 14.2(4) & 138.2 & $\pm$ 1.3 & 25.0 & 1.31(4) & 0.9 & $\pm$ 1.6 &  \\
$f_3$ & 8.6047 & (3) & 18.0(3) & 194.5 & $\pm$ 1.1 & 28.7 & 11.6(3) & 200.5 & $\pm$ 1.8 & 15.2 & 1.56(5) & -5.9 & $\pm$ 2.1 &  \\
$2f_1$ & 13.4568 & & 11.3(4) & 115.6 & $\pm$ 1.9 & 25.8 & 9.0(3) & 118.1 & $\pm$ 1.9 & 18.8 & 1.25(6) & -3 & $\pm$ 3 &  \\
$f_0+f_2$ & 13.5385 & & 8.8(4) & 40.5 & $\pm$ 2.3 & 20.2 & 6.8(3) & 46.8 & $\pm$ 2.8 & 14.5 & 1.29(8) & -6 & $\pm$ 4 &  \\
$2f_0$ & 10.2823 & & 10.0(4) & 311.3 & $\pm$ 2.0 & 16.2 & 7.8(3) & 317.7 & $\pm$ 2.6 & 11.7 & 1.28(7) & -6 & $\pm$ 3 &  \\
$f_1+f_2$ & 15.1258 & & 6.4(3) & 122.7 & $\pm$ 3.0 & 13.3 & 6.1(3) & 132.1 & $\pm$ 2.9 & 12.9 & 1.05(8) & -9 & $\pm$ 4 &  \\
$f_2-f_1$ & 1.6690 & & 9.5(4) & 155.3 & $\pm$ 2.0 & 9.5 &  &  & &  &  &  &  \\
$f_1-f_0$ & 1.5872 & & 8.8(3) & 106.2 & $\pm$ 2.1 & 8.6 & 8.3(3) & 84.2 & $\pm$ 2.4 & 4.1 & 1.06(5) & 22 & $\pm$ 3 &  \\
$2f_1+f_2$ & 21.8541 & & 3.3(3) & 274.9 & $\pm$ 6.0 & 6.7 & 2.9(3) & 270.7 & $\pm$ 6.1 & 7.3 & 1.1(2) & 4 & $\pm$ 9 &  \\
$f_1+f_3$ & 15.3331 & & 3.1(3) & 61.6 & $\pm$ 7.1 & 6.1 &  &  & &  &  &  &  \\
$2f_1-f_0$ & 8.3156 & & 3.6(4) & 289.8 & $\pm$ 4.9 & 5.7 &  &  & &  &  &  &  \\
$3f_1$ & 20.1851 & & 3.0(3) & 348.8 & $\pm$ 6.9 & 5.7 & 2.6(3) & 355.3 & $\pm$ 6.8 & 6.4 & 1.2(2) & -6 & $\pm$ 10 &  \\
$f_4$ & 8.9793 & (3) & 3.4(3) & 349.4 & $\pm$ 6.0 & 5.4 &  &  & &  &  &  &  \\
$f_0+f_3$ & 13.7459 & & 2.3(4) & 219.1 & $\pm$ 7.7 & 5.4 & 1.9(3) & 245 & $\pm$ 11 & 4.1 & 1.2(3) & -26 & $\pm$ 13 &  \\
$f_2-f_0$ & 3.2562 & & 5.3(3) & 298.7 & $\pm$ 3.8 & 5.3 &  &  & &  &  &  &  \\
$2f_0+f_1$ & 17.0107 & & 2.8(3) & 92.8 & $\pm$ 7.3 & 4.5 &  &  & &  &  &  &  \\
\hline
\end{tabular}
\end{center}
\end{table*}

%

\section{Discussion}

\begin{figure}
\centering
\includegraphics[width=9cm]{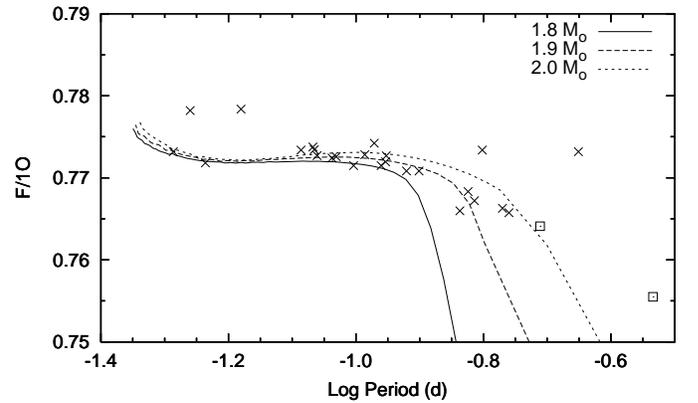}
\caption{Theoretically calculated Petersen diagram for the frequency ratio between the fundamental mode and first overtone for solar metallicity and three different star masses.
Observed frequency ratios of Galactic HADS are plotted as crosses.  The $F/1O$ frequency ratios for GSC~762-110 and V829~Aql are plotted as squares.
Figure adapted from Poretti et al. (\cite{hads}).}
\label{track}
\end{figure}

GSC~762-110 is the first known triple-mode radial pulsator with a fundamental period within the traditional $\delta$ Scuti regime.
Its period of 0.1945~days is shorter than the known double-mode HADS \object{VX~Hya} (0.2234~days; Fitch, \cite{vxhya}).
We find a frequency ratio $F/1O = 0.7641$ between the fundamental mode and the first overtone and $1O/2O = 0.8012$
between the first and second overtone. 
From evolutionary models theoretical pulsation constants can be derived starting from various initial conditions (see e.g. Fitch, \cite{fitch}).
Poretti et al. (\cite{hads}) give Petersen diagrams for the $F/1O$ ratio for different masses and metallicities.
There is very good agreement in the case of GSC~762-110 with the $2M_{\sun}$ track for solar metallicity. 
This can be seen in Petersen diagram in Fig.~\ref{track} (adapted from Poretti et al., \cite{hads}).  
The figure also includes V829~Aql which is likely a $2.1M_{\sun}$ HADS (Handler et al., \cite{v829aql}).
Similar agreement is found for GSC~762-110 with the theoretical ratios for the $2M_{\sun}$ models calculated by Fitch (\cite{fitch}) for both ratios,
although these suggest a somewhat larger value for the fundamental period.
An overview of the frequency ratios for the known Galactic triple-mode pulsators is given in Table~\ref{triple}: listed are the fundamental period in days, the two frequency ratios,
and the amplitude and the amplitude ratios in $V$ for the overtone modes compared to the fundamental mode. 

\begin{table*}
\begin{center}
\caption{Comparison of the characteristics of Galactic triple-mode variables.}
\label{triple}
\begin{tabular}{ccccc}
\hline
 & AC~And & V823~Cas & V829~Aql & GSC~762-110 \\
\hline
Reference & Fitch \& Szeidl, \cite{acand1} & Jurcsik et al., \cite{v823cas} & Handler et al., \cite{v829aql} & this paper \\
Period F (d) & 0.7112 & 0.6690 & 0.2924 & 0.1945 \\
$F/1O$  & 0.7383 & 0.7663 & 0.7555 & 0.7641 \\
$1O/2O$ & 0.8018 & 0.8015 & 0.7997 & 0.8012 \\
$A_F$ (mmag) & 204 & 86 & 82 & 75 \\
$A_{1O}/A_F$ & 0.84 & 1.51 & 1.04 & 1.00 \\
$A_{2O}/A_F$ & 0.35 & 0.26 & 0.35 & 0.53 \\
\hline
\end{tabular}
\end{center}
\end{table*}

Because the fundamental period of the double-mode HADS VX~Hya falls between those of GSC~762-110 and V829~Aql one might wonder
whether it also is a triple-mode pulsator.  
However the ASAS-3 data of VX~Hya do not reveal any excited mode with an amplitude above 0.01 mag in the expected frequency range.

The remaining two pulsation modes $f_3$ and $f_4$ in GSC~762-110 can only be non-radial pulsation modes, 
when comparing these with the tables from Fitch (\cite{fitch}), as the ratios with the radial modes are too high.
Not much theoretical research has been done on non-radial modes in radial pulsators, 
although they have been found to be excited in some double-mode HADS as well, e.g. \object{RV~Ari} (P\'ocs et al., \cite{rvari}),
and recently also in the double-mode RR~Lyrae star \object{AQ~Leo} (Gruberbauer et al., \cite{aqleo}).
In addition there are indications of a non-radial $\gamma$~Dor mode in V823~Cas (Jurcsik et al., \cite{v823cas}).

As seen from Table~\ref{freq}, the amplitude ratio $A_B/A_V$ for $f_3$ is fairly large, about 25\% larger than the amplitude ratios for the radial modes.  
Both this amplitude ratio and the negative phase difference are important constraints to determine
the pulsation mode (see for example Fig.~5 from Daszy\'nska-Daszkiewicz, \cite{dasz}).
The amplitude of $f_3$ looks about halved when comparing our $V$ data (from the end of 2006 and early 2007) with those of ASAS-3 (data gathered between 2003 to 2006), 
while the amplitudes of the radial modes have remained stable.  
In addition the $f_3$ frequency itself seems somewhat smaller from our data.  
Again the frequencies of the radial modes have remained stable.

The last independent frequency $f_4$ is even more remarkable.  It has been found in ASAS-3 $V$ data with a fairly large amplitude, 
and in our $B$ data with a smaller amplitude.  Surprisingly it was not found at all in our $V$ data, 
although other frequencies were detected with a significantly smaller amplitude in $V$ than the amplitude in the ASAS-3 data.  
Also there seems to be a significant difference between the value of the frequency itself as found from ASAS-3 and from our $B$ data.
The amplitude of this frequency must have diminished considerably over a short time, and its value must have shifted as well.  
Unfortunately there is not enough ASAS-3 data available for the individual years to investigate any changes of $f_3$ and $f_4$ during the years 2003 to 2006.

Further studies of GSC~762-110 are encouraged, also in view of possible period changes that have been found in AC~And and V823~Cas (Jurcsik et al., \cite{v823cas}).
It should also be interesting to see whether the non-radial frequencies return more prominently in the future.  

\begin{acknowledgements}
Use of NASA's Astrophysics Data System, and the SIMBAD and VizieR 
databases operated at the Centre de Donn\'ees Astronomiques (Strasbourg) in France
is gratefully acknowledged.
\end{acknowledgements}

\end{document}